\def\be{\begin{equation}}
\def\ee{\end{equation}}
\def\ba{\begin{array}}
\def\ea{\end{array}}

\def\Nb{{I\!\! N}}

\def\Cb{\ \hbox{\vrule width 0.6pt height 6pt depth 0pt
              \hskip -3.2 pt} C}
\documentstyle[12pt]{article}
\begin{document}
\baselineskip=22pt \setcounter{page}{1} \centerline{\Large\bf
Entanglement of Formation for a Class} \vspace{2ex}
\centerline{\Large\bf of Quantum States} \vspace{5ex}
\begin{center}
Shao-Ming Fei$^{\ddag\dag}$\footnote{Email: fei@uni-bonn.de},
J\"urgen Jost$^{\dag}$\footnote{Email: jjost@mis.mpg.de}, Xianqing
Li-Jost$^{\dag}$\footnote{Email:  Xianqing.Li-Jost@mis.mpg.de}
and Guo-Fang Wang$^{\dag}$\footnote{Email:  Guofang.Wang@mis.mpg.de}
\end{center}
\bigskip

\begin{center}
\begin{minipage}{5.2in}
\parskip=6pt

$^\ddag$Department of Mathematics, Capital Normal University,
Beijing 100037

$^\dag$Max-Planck-Institute for Mathematics in the Sciences, 04103
Leipzig

$^\ddag$Institut f{\"u}r Angewandte Mathematik, Universit{\"a}t
Bonn, 53115 Bonn

\end{minipage}
\end{center}

\vskip 2 true cm
\parindent=18pt
\parskip=6pt
\begin{center}
\begin{minipage}{5in}
\vspace{3ex} \centerline{\large Abstract} \vspace{4ex}
Entanglement of formation for a class of higher dimensional
quantum mixed states is studied in terms of a generalized formula
of concurrence for $N$-dimensional quantum systems. As
applications, the entanglement of formation for a class of
$16\times 16$ density matrices are calculated.

\bigskip
\medskip
\bigskip

\vfill


\end{minipage}
\end{center}

\bigskip
Quantum entanglement plays important roles in quantum
communication, information processing and quantum computing
\cite{DiVincenzo}, such as in the investigation of quantum
teleportation \cite{teleport,teleport1,teleport2}, dense coding
\cite{dense}, decoherence in quantum computers \cite{DiVincenzo}
and the evaluation of quantum cryptographic schemes \cite{crypto}.
To quantify entanglement, a well justified and mathematically
tractable measure of entanglement is needed. A number of
entanglement measures such as the entanglement of formation and
distillation \cite{Bennett96a,BBPS,Vedral}, negativity
\cite{Peres96a,Zyczkowski98a}, and relative entropy
\cite{Vedral,sw} have been proposed for bipartite states
[6,8,12-17]. Nevertheless, most proposed measures of entanglement
involve extremizations which are difficult to handle analytically.

The ``entanglement of formation'' is intended to quantify the
amount of quantum communication required to create a given state
\cite{Bennett96a}. Although it is defined for arbitrary dimensions,
so far no explicit analytic formulae for entanglement of formation
have been found for systems larger than a pair of qubits, \
due to the fact that two dimensional
bipartite mixed states are special in many ways \cite{ww}, except
for some special symmetric states \cite{th}.

In this letter we study the entanglement of formation for a class
of higher dimensional quantum mixed states. For certain
$N$-dimensional pure quantum systems, we show that the
entanglement of formation is a monotonically increasing function
of a kind of generalized concurrence. As applications, the
entanglement of formation for a class of $16\times 16$ density
matrices is calculated in detail. The method applies to a large
class of quantum states. The construction of these states are
presented for $N$ dimensional, $N=2^{k+1}$, $2\leq k \in\Nb$,
bipartite systems.

Let ${\cal H}$ be an $N$-dimensional complex Hilbert space with
orthonormal basis $e_i$, $i=1,...,N$.
A pure state on ${\cal H}\otimes{\cal H}$ is generally of the form,
\begin{equation}\label{psi}
\vert\psi\rangle=\sum_{i,j=1}^N a_{ij}e_i\otimes e_j,~~~~~~a_{ij}\in\Cb
\end{equation}
with normalization
\begin{equation}\label{norm}
\sum_{i,j=1}^N a_{ij}a_{ij}^\ast=1\,.
\end{equation}
The entanglement of formation $E$ is defined as the entropy of
either of the two sub-Hilbert space ${\cal H}\otimes{\cal H}$
\cite{BBPS},
\be\label{epsi} E(|\psi \rangle) = - {\mbox{Tr}\,}
(\rho_1 \log_2 \rho_1) = - {\mbox{Tr}\,} (\rho_2 \log_2 \rho_2)\,,
\ee where $\rho_1$ (resp. $\rho_2$) is the partial trace of $\bf
|\psi\rangle\langle\psi|$ over the first (resp. second) Hilbert
space of ${\cal H}\otimes{\cal H}$.

Let $A$ denote the matrix with
entries given by $a_{ij}$ in (\ref{psi}). $\rho_1$ can be
expressed as \be \rho_1=AA^\dag. \ee
For a given density matrix of a pair of quantum systems on
${\cal H}\otimes{\cal H}$, consider all possible pure-state decompositions of
$\rho$, i.e., all ensembles of state $|\psi_i \rangle$ of the form (\ref{psi})
with probabilities $p_i$,
$$
\rho = \sum_{i=1}^M p_i |\psi_i \rangle \langle\psi_i|,~~~~\sum_{i=1}^M p_i =1
$$
for some $M\in\Nb$. The entanglement of formation for the mixed
state $\rho$ is defined as the average entanglement of the pure
states of the decomposition, minimized over all possible
decompositions of $\rho$,
\be\label{erho} E(\rho) = \mbox{min}\,
\sum_{i=1}^M p_i E(|\psi_i \rangle).
\ee

It is a challenge to calculate (\ref{erho}) for general $N$. Till
now a general explicit formula of $E(\rho)$ is obtained only for
the case $N=2$. In this case (\ref{epsi}) can be written as
$$
E(|\psi \rangle)|_{N=2} =h(\frac{1+\sqrt{1-C^2}}{2}),
$$
where
$$
h(x) = -x\log_2 x - (1-x)\log_2 (1-x),
$$
$C$ is called concurrence \cite{HillWootters}:
$$
C(|\psi \rangle) = |\langle \psi | \tilde{\psi} \rangle |
=2|a_{11}a_{22}-a_{12}a_{21}\vert,
$$
where $|\tilde{\psi}\rangle = \sigma_y\otimes\sigma_y |\psi^* \rangle$,
$|\psi^* \rangle$ is the complex conjugate of $|\psi\rangle$, $\sigma_y$
is the Pauli matrix, $\sigma_y= \left(
\begin{array}{cc}0&-i\\
i&0\end{array}\right)$.

As $E$ is a monotonically increasing function of $C$, $C$ can be
also taken as a kind of measure of entanglement. Calculating
(\ref{erho}) is reduced to calculate the corresponding minimum of
$C(\rho) = \mbox{min}\, \sum_{i=1}^M p_i C(|\psi_i \rangle)$,
which simplifies the problems.

For $N\geq 3$, there is no such concurrence $C$ in general. The
concurrences discussed in \cite{concu} can be only used to judge
whether a pure state is separable (or maximally entangled) or not
\cite{qsep,separ3}. The entanglement of formation is no longer a
monotonically increasing function of these concurrences.
Nevertheless, for a special class of quantum states, we can find
certain quantities (generalized concurrence) to simplify the
calculation of the corresponding entanglement of formation.

{\sf [Theorem 1]}. If $AA^\dag$ has only two non-zero eigenvalues
(each of which may be degenerate), the maximal non-zero diagonal
determinant $D$ of $AA^\dag$ is a generalized concurrence. The
entanglement of formation of the corresponding pure state is a
monotonically increasing function of $D$.

{\sf [Proof]}. Let $\lambda_1$ (resp. $\lambda_2$) be the two non-zero
eigenvalues of $AA^\dag$ with degeneracy $n$ (resp. $m$), $n+m\leq N$.
That is,
\be\label{D}
D=\lambda_1^n\lambda_2^m\,.
\ee
From the normalization of $|\psi\rangle$, one has $Tr (AA^\dag) =1$, i.e.,
\be\label{nm}
n\lambda_1+m\lambda_2 =1\,.
\ee
$\lambda_1$ (resp. $\lambda_2$) takes values $(0,\frac{1}{n})$ (resp. $(0,\frac{1}{m})$).
In this case the entanglement of formation of $|\psi \rangle$ is given by
\be\label{enm}
E(|\psi \rangle)=-n \lambda_1 \log_2 \lambda_1 -m \lambda_2 \log_2 \lambda_2\,.
\ee
According to (\ref{D}) and (\ref{nm}) we get
\be\label{ded}
\frac{\partial E}{\partial D}=
\frac{m\lambda_1^{1-n}}{1-n\lambda_1-m\lambda_1}
\left(\frac{1-n\lambda_1}{m}\right)^{1-m}
\log_2 \frac{1-n\lambda_1}{m\lambda_1}\,,
\ee
which is positive for $\lambda_1\in(0,\frac{1}{n})$. Therefore $E(|\psi
\rangle)$ is a monotonically increasing function of $D$. $D$ is a
generalized concurrence and
can be taken as a kind of measure
of entanglement in this case. \hfill $\rule{2mm}{2mm}$

{\sf Remark:} We have assumed that $\lambda_1$, $\lambda_2\neq 0$
in our theorem. In fact the right hand side of (\ref{ded}) keeps
positive even when $\lambda_1$ (or equivalently $\lambda_2$) goes
to zero. Hence $E(|\psi\rangle)$ is a monotonically increasing
function of $D$ for $\lambda_1\in [0,\frac{1}{n}]$ (resp.
$\lambda_2\in [0,\frac{1}{m}]$) satisfying the relation
(\ref{nm}). Nevertheless if $\lambda_1=0$ (or $\lambda_2=0$), from
(\ref{D}) one gets $D=0$, which does not necessarily mean that the
corresponding state $|\psi \rangle$ is separable. As
$E(|\psi\rangle)$ is just a monotonically increasing function of
$D$, $D$ only characterizes the relative degree of the
entanglement among the class of these states.

From (\ref{nm}) and (\ref{enm}), the quantum states with the measure of
entanglement characterized by
$D$ are generally entangled. They are separated when $n=1$,
$\lambda_1\to 1$ ($\lambda_2\to 0$) or $m=1$, $\lambda_2\to 1$ ($\lambda_1\to 0$).
For the case $n=m>1$, all the pure states in this class are non-separable.
In this case, we have
\be\label{epsinn}
E(|\psi \rangle)=n \left(-x\log_2 x
- (\frac{1}{n}-x)\log_2 (\frac{1}{n}-x)\right),
\ee
where
$$
x = \frac{1}{2}\left(\frac{1}{n}+\sqrt{\frac{1}{n^2}(1-d^2)}\right)
$$
and
\be\label{GC}
d\equiv 2nD^{\frac{1}{2n}}=2n\sqrt{\lambda_1\lambda_2}.
\ee
In this case we define $d$ to be the generalized concurrence. $d$ takes value from
$0$ to $1$. From (\ref{epsinn}) one can show that $E(d)$ is a convex function
(that is, curving upward):
$$
\frac{\partial^2 E}{\partial d^2}=\frac{\log \frac{1+\sqrt{1-d^2}}
{1-\sqrt{1-d^2}}-2\sqrt{1-d^2}}{(1-d^2)^{3/2}\log 4}>0,~~~~~\forall~ d\in [0,1]\,.
$$

Instead of calculating $E(\rho)$ directly, one may calculate the
minimum decomposition of $D(\rho)$ or $d(\rho)$ to simplify the
calculations. In the following, as an example, we calculate the
entanglement of formation for a class of mixed states with $N=4$.

We consider a class of pure states (\ref{psi}) with the matrix $A$ given by
\be\label{a} A=\left( \ba{cccc}
0&b&a_1&b_1\\
-b&0&c_1&d_1\\
a_1&c_1&0&-e\\
b_1&d_1&e&0 \ea \right),
\ee
$a_1,b_1,c_1,d_1,b,e\in\Cb$. The
matrix $AA^\dag$ has two eigenvalues with degeneracy two, i.e.,
$n=m=2$.
\be\label{detaa}
det(AA^\dag)=|b_1c_1-a_1d_1+be|^4.
\ee
According to our theorem, the generalized concurrence
\be\label{d1}
d=4|b_1c_1-a_1d_1+be|
\ee
is a kind of measure of
entanglement for all pure states of the form (\ref{a}).

Let $p$ be a $16\times 16$ matrix with only non-zero entries
$p_{1,16}=p_{2,15}=-p_{3,14}=p_{4,10}=p_{5,12}=p_{6,11}
=p_{7,13}=-p_{8,8}=-p_{9,9}=p_{10,4}=p_{11,6}=p_{12,5}
=p_{13,7}=-p_{14,3}=p_{15,2}=p_{16,1}=1$.
$d$ in (\ref{d1}) can be written as
\be\label{dp}
d=|\langle \psi |p\psi^* \rangle |\equiv|\langle\langle \psi |\psi \rangle\rangle |,
\ee
where $\langle\langle \psi |\psi \rangle\rangle=
\langle \psi |p\psi^* \rangle$.

We now calculate the entanglement of formation for a special class of mixed states.
Let $\Psi$ denote the set of pure states (\ref{psi})
with $A$ given as the form of (\ref{a}). We consider all mixed states with
density matrix $\rho$ such that its decompositions are of the form
\be\label{rho1}
\rho = \sum_{i=1}^M p_i |\psi_i \rangle \langle \psi_i|,~~~~\sum_{i=1}^M
p_i =1,~~~~|\psi_i\rangle\in\Psi.
\ee

Let $s\leq 16$ be the rank of $\rho$ and $|v_i\rangle$, $i=1,...,s$, be
a complete set of orthogonal eigenvectors
corresponding to the nonzero eigenvalues of $\rho$, such that
$\langle v_i|v_i \rangle$ is equal to the $i$th eigenvalue.
Other decomposition $\{ |w_i\rangle \}$ of $\rho$
can then be obtained through unitary transformations:
\begin{equation}\label{w}
| w_i \rangle = \sum_{j=1}^s U_{ij}^* | v_j\rangle,
\end{equation}
where $U$ is a $t \times t$ unitary matrix, $t\geq s$.
The states $|w_i\rangle$ are so normalized that
$\rho = \sum_i |w_i\rangle \langle w_i|$. It is obvious that for any
$|\psi_i\rangle\in\Psi$, the state of complex
linear combination of $|\psi_i\rangle$
(unitary transformations) also belongs to $\Psi$.

The decomposition according to the orthogonal eigenvectors $|v_i\rangle$ of $\rho$
is not the one satisfying (\ref{erho}) in general.
As the generalized concurrence can be written as the form (\ref{dp}) for
a pure state, we consider the quantity
$\langle\langle w_i|{w}_j\rangle\rangle$. From (\ref{w}) we have
$$
\langle\langle w_i|{w}_j\rangle\rangle =(U \tau U^T)_{ij},
$$
where the matrix $\tau$ is defined by
$\tau_{ij} \equiv \langle\langle v_i|v_j \rangle\rangle$. The matrix $p$ in
(\ref{dp}) is a symmetric one. Therefore $\tau$
is also symmetric and can always be diagonalized by a unitary matrix $U$
such that $U \tau U^T=diag (\Lambda_1,...,\Lambda_s)$ \cite{HJ}.
The diagonal elements $\Lambda_i$, in deceasing order,
can always be made to be real and non-negative.
Since $U \tau \tau^{*} U^{\dag}$ is also diagonal, $\Lambda_i$ are just the
square roots of the eigenvalues of $\tau \tau^{*}$.
It is straightforward to check that they are also the
eigenvalues of the Hermitian
matrix $R \equiv \sqrt{\sqrt{\rho}p{\rho^\ast}p\sqrt{\rho}}$,
or, alternatively, the square roots of the eigenvalues of the
non-Hermitian matrix $\rho p{\rho^\ast}p$.

Hence there always exits a decomposition consisting of states
$|w_i\rangle$, $i=1,\ldots ,s$, such that
\begin{equation}\label{ww}
\langle\langle w_i|{w}_j\rangle\rangle = \Lambda_i\delta_{ij}.
\end{equation}
We can now deal with the problem in a way similar to \cite{HillWootters}.
Set
\begin{equation}
|y_1\rangle = |w_1\rangle,~~~~
|y_j\rangle = i|w_j\rangle~~~~ { \rm for } ~ j = 2 , ... , s.
\end{equation}
Any decomposition can be written
in terms of the states $|y_i\rangle$ via the equation
$$
| z_i \rangle = \sum_{j=1}^s V_{ij}^* | y_j\rangle,
$$
where $V$ is a $t \times s$ matrix whose $s$ columns are
othonormal vectors.

The average concurrence
of a general decomposition is given by
\begin{equation}
\langle d\rangle = \sum_i |(VYV^T)_{ii}|=\sum_i \Bigl| \sum_j
(V_{ij})^2 Y_{jj} \Bigr|,
\end{equation}
where $Y$ is the real diagonal matrix defined by
$Y_{ij} = \langle\langle y_i|y_j \rangle\rangle$. Using the fact
that $\sum_i |(V_{ij})^2| = 1$, one gets
$$
\langle d\rangle \ge | \sum_{ij} (V_{ij})^2 Y_{jj} |
\ge \Lambda_1 - \sum_{i=2}^{16}\Lambda_i.
$$

Therefore the minimum decomposition of the generalized concurrence
is
\be
\label{drho} d(\rho)=\Lambda_1 - \sum_{i=2}^{16}\Lambda_i.
\ee
Similar to the case $N=2$, there are decompositions such that the
generalized concurrence of each individual state is equal to
$d(\rho)$. Therefore the average entanglement is $E(d(\rho))$.

Different from the case $N=2$, the entanglement of formation of
density matrices (\ref{rho1}) can not be zero in general. As every
individual pure state in the decompositions is generally an entangled
one, this class of mixed states are not separable.

In the following we call an $N$-dimensional pure state (\ref{psi})
\underline{$d$-computable} if $A$ satisfies the following relations:
\be\label{1} \ba{l}
det(AA^\dag) = ([ A ][ A ]^\ast)^{N/2}\\[3mm]
det(AA^\dag - \lambda Id_N) =(\lambda^2 - \| A \| \lambda + [ A ][ A ]^\ast)^{N/2},
\ea
\ee
where $[A]$ and $\|A \|$ are any quadratic forms of $a_{ij}$
(these quadratic forms could be different for different matrix $A$),
$Id_N$ is the $N\times N$
identity matrix. We denote ${\cal A}$ the set of matrices
satisfying (\ref{1}), which implies that for $A\in{\cal A}$, $AA^\dag$ has at most
two different eigenvalues and each one has order $N/2$.
Formula (\ref{drho}) can be generalized to general $N^2\times N^2$
density matrices with decompositions on $N$-dimensional
$d$-computable pure states.

A class of $N$-dimensional, $N = 2^k$, $2\leq k\in\Nb$,
$d$-computable states has been constructed in \cite{a8}.
These states give rise to a special class of
density matrices with decompositions in these pure states,
and the entanglement of formation for these density matrices can
be calculated analytically according to the method above.

Let $A$ be an $N\times N$ matrix with entries $a_{ij}\in
\Cb$, $i,j=1,...,N$, with the following properties:

Set
$$
A_2= \left(
\begin{array}{cc}
a&-c \\[3mm]
c&d\\[3mm]
\end{array}
\right),
$$
where $a,c,d \in \Cb$.
For any $b_1,c_1 \in \Cb$, a $4\times 4$ matrix $A_4\in{\cal A}$ can
be constructed in the following way,
\be\label{ha4}
A_4= \left(
\begin{array}{cc}
B_2&A_2\\[3mm]
-A_2^t&C_2^t\\[3mm]
\end{array}
\right) = \left(
\begin{array}{cccc}
0&b_1&a&-c\\[3mm]
-b_1&0&c&d\\[3mm]
-a&-c&0&-c_1\\[3mm]
c&-d&c_1&0
\end{array}
\right),
\ee
where
$$
B_2 = b_1J_2, ~~~~ C_2 = c_1J_2, ~~~ J_2= \left(
\begin{array}{cc}
0&1 \\[3mm]
-1&0\\[3mm]
\end{array}
\right).
$$
$A_4$ satisfies the relations in (\ref{1}):
$$
\begin{array}{l}
\left| A_4 A^\dag_4 \right|=[(b_1c_1+a d + c^2)(b_1c_1+a d + c^2)^\ast]^2=
([ A_4 ][ A_4 ]^\ast)^2,\\[3mm]
\left| A_4 A^\dag_4 - \lambda Id_4 \right| = (\lambda^2 -
(b_1b_1^\ast+c_1c_1^\ast+aa^\ast+2cc^\ast+dd^\ast)\lambda\\[3mm]
~~~~~~~~~~~~~~~~~~~~~~+ (b_1c_1+ a d + c^2)(b_1c_1+ a d + c^2)^\ast)^2\\[3mm]
~~~~~~~~~~~~~~~~~~~= (\lambda^2 - \| A_4 \|\lambda + [ A_4 ][ A_4 ]^\ast)^2,
\end{array}
$$
where
\be
[ A_4 ]=(b_1c_1+a d + c^2),~~~
\| A_4 \|=b_1b_1^\ast+c_1c_1^\ast+aa^\ast+2cc^\ast+dd^\ast.
\ee

$A_8\in{\cal A}$ can be obtained from $A_4$,
\be\label{a8} A_8=
\left(
\begin{array}{cc}
B_4&A_4\\[3mm]
-A_4^t&C_4^t\\[3mm]
\end{array}
\right), \ee where
\be\label{i4} B_4 = b_2J_4, ~~~~C_4 = c_2J_4,
~~~~ J_4= \left(
\begin{array}{cccc}
0&0&0&1\\[3mm]
0&0&1&0\\[3mm]
0&-1&0&0\\[3mm]
-1&0&0&0
\end{array}
\right),~~~~
b_2,~c_2 \in \Cb.
\ee

For general construction of high dimensional matrices
$A_{2^{k+1}}\in{\cal A}$, $2 \leq k\in\Nb$, we have
\be\label{a2k}
A_{2^{k+1}}= \left(
\begin{array}{cc}
B_{2^{k}}&A_{2^{k}}\\[3mm]
(-1)^{\frac{k(k+1)}{2}}A_{2^{k}}^t&C_{2^{k}}^t
\end{array}
\right) \equiv\left(
\begin{array}{cc}
b_{k}J_{2^{k}}&A_{2^{k}}\\[3mm]
(-1)^{\frac{k(k+1)}{2}}A_{2^{k}}^t&c_{k}J_{2^{k}}^t
\end{array}
\right),
\ee

\be
\label{i2k}
J_{2^{k+1}}= \left(
\begin{array}{cc}
0&J_{2^{k}}\\[3mm]
(-1)^{\frac{(k+1)(k+2)}{2}}J_{2^{k}}^t&0\\[3mm]
\end{array}
\right),
\ee
where $b_k, c_k \in \Cb$,
$B_{2^{k}}=b_{k}J_{2^{k}}$, $C_{2^{k}}=c_{k}J_{2^{k}}$.
It can be verified that
$A_{2^{k}}$ satisfies the following relations \cite{a8}:
\be\label{thm2}
\ba{l}
|A_{2^{k+1}}A_{2^{k+1}}^\dag|=([A_{2^{k+1}}][A_{2^{k+1}}]^*)^{2^k}
=[((-1)^{\frac{k(k+1)}{2}}b_kc_k-[A_{2^{k}}])
((-1)^{\frac{k(k+1)}{2}}b^*_kc^*_k-[A_{2^{k}}]^*)]^{2^k},\\[3mm]
|A_{2^{k+1}}A_{2^{k+1}}^\dag-\lambda
Id_{2^{k+1}} |
=(\lambda^2-||A_{2^{k+1}}||\lambda+[A_{2^{k+1}}][A_{2^{k+1}}]^*)^{2^k}.
\ea
\ee
Therefore the states given by (\ref{a2k}) are $d$-computable.
In terms of (\ref{GC}) the generalized concurrence for these states is given by
$$
d_{2^{k+1}}=2^{k+1}\vert[A_{2^{k+1}}]\vert=2^{k+1}\vert b_kc_k+
b_{k-1}c_{k-1}+...+b_1c_1+ad+c^2\vert.
$$

Let $p_{2^{k+1}}$ be a symmetric anti-diagonal $2^{2k+2}\times 2^{2k+2}$ matrix with
all the anti-diagonal elements $1$ except for those
at rows $2^{k+1}-1 + s(2^{k+2}-2)$, $2^{k+1} + s(2^{k+2}-2)$,
$2^{k+2}-1 + s(2^{k+2}-2)$, $2^{k+2} + s(2^{k+2}-2)$,
$s=0,...,2^{k+1}-1$, which are $-1$. $d_{2^{k+1}}$ can be written as
\be\label{dkp}
d_{2^{k+1}}=|\langle \psi_{2^{k+1}} |p_{2^{k+1}}\psi_{2^{k+1}}^{*} \rangle |
\equiv|\langle\langle \psi_{2^{k+1}} |\psi_{2^{k+1}} \rangle\rangle |,
\ee
where
\be\label{psi2k1}
\vert\psi_{2^{k+1}}\rangle=\sum_{i,j=1}^{2^{k+1}} (A_{2^{k+1}})_{ij}\,e_i\otimes e_j.
\ee
According to the calculations on entanglement of formation for $d$-computable states,
for a $2^{2k+2}\times 2^{2k+2}$ density matrix $\rho_{2^{k+2}}$ with
decompositions on pure states of the form (\ref{psi2k1}), its entanglement of formation
is given by $E(d_{2^{k+1}}(\rho_{2^{2k+2}}))$, where
\be\label{d2k1}
d_{2^{k+1}}(\rho_{2^{2k+2}})=\Omega_1 - \sum_{i=2}^{2^{2k+2}}\Omega_i,
\ee
and $\Omega_i$, in decreasing order, are the
the square roots of the eigenvalues of the
matrix $\rho_{2^{2k+2}} p_{2^{k+1}}{\rho_{2^{2k+2}}^\ast}p_{2^{k+1}}$.

We have studied the entanglement of formation for a class
of higher dimensional quantum mixed states. It is shown that for certain
$N$-dimensional pure quantum systems, the
entanglement of formation is a monotonically increasing function
of a generalized concurrence. From this generalized concurrence
the entanglement of formation for
a large class of quantum states can be calculated analytically.
The physical properties of these states are remained to be studied further.


\bigskip
\medskip
\begin{thebibliography}{99}

\bibitem{DiVincenzo} See, for example, D.P. DiVincenzo,
{\em Science} {\bf 270}, 255 (1995).

\bibitem{teleport} C.H. Bennett, G. Brassard, C. Cr\'epeau,
       R. Jozsa, A. Peres,
       and W.K. Wootters, {\em Phys. Rev. Lett.} {\bf 70}, 1895 (1993).

\bibitem{teleport1} S. Albeverio and S.M. Fei,
          {\em Phys. Lett.} {\bf A 276}, 8-11(2000).

\bibitem{teleport2} S. Albeverio and S.M. Fei and W.L. Yang, {\em Commun. Theor.
          Phys.} {\bf 38}, 301-304(2002); {\em Phys. Rev.} {\bf A 66}, 012301(2002).

\bibitem{dense} C.H. Bennett and S.J. Wiesner,
        {\em Phys. Rev. Lett.} {\bf 69}, 2881 (1992).

\bibitem{crypto} See, for example, C.A. Fuchs, N. Gisin,
         R.B. Griffiths, C-S. Niu, and
         A. Peres, {\em Phys. Rev.}, {\bf A 56}, 1163 (1997) and
         references therein.

\bibitem{Bennett96a}
C.H. Bennett, D.P. DiVincenzo, J.A. Smolin, and W.K. Wootters,
Phys. Rev. {\bf A 54},  3824  (1996).

\bibitem{BBPS} C.H. Bennett, H.J. Bernstein, S. Popescu,
and B. Schumacher, {\em Phys.~Rev.~A} {\bf 53}, 2046 (1996).

\bibitem{Vedral} V.~Vedral, M.B.~Plenio, M.A.~Rippin, and
P.L.~Knight, {\em Phys. Rev. Lett.} {\bf 78}, 2275 (1997);\\
V.~Vedral, M.B.~Plenio, K.~Jacobs, and P.L.~Knight, {\em Phys. Rev. A}
{\bf 56}, 4452 (1997);\\
V.~Vedral and M.B.~Plenio, {\em Phys. Rev. A} {\bf 57}, 1619 (1998).

\bibitem{Peres96a}
A. Peres, Phys. Rev. Lett. {\bf 77} 1413 (1996).

\bibitem{Zyczkowski98a}
K. \.Zyczkowski and P. Horodecki, Phys. Rev. A {\bf 58},  883  (1998).

\bibitem{sw}
B. Schumacher and M.D. Westmoreland, {\it Relative entropy in
quantum information theory}, quant-ph/0004045.

\bibitem{Horodecki} M.~Horodecki, P.~Horodecki, and R.~Horodecki,
{\em Phys. Rev. Lett.} {\bf 80}, 5239 (1998).

\bibitem{Rains} E.M.~Rains,
{\em IEEE Trans. Inform. Theory} {\bf 47}, 2921-2933 (2001).

\bibitem{HillWootters} S.~Hill and W.K.~Wootters, {\em Phys. Rev. Lett.}
{\bf 78}, 5022 (1997).\\
W.K.~Wootters, {\em Phys. Rev. Lett.} {\bf 80}, 2245 (1998).

\bibitem{ww}
R.F. Werner and M.M. Wolf, {\em Phys. Rev.} {\bf A 61}, 062102
(2000).

\bibitem{th}
B.M. Terhal, K. Gerd and K.G.H. Vollbrecht, {Phys. Rev. Lett.} {\bf 85},
2625 (2000).

\bibitem{concu}
A.Uhlmann, {\em Phys. Rev.} {\bf A 62}, 032307 (2000).\\
S. Albererio and S.M. Fei, {\em J. Opt. B: Quantum Semiclass. Opt.}
{\bf 3} 1-5(2001).\\
P. Rungta, V. Bu$\check{\rm z}$ek, C.M. Caves, M. Hillery, G.J.
Milburn, {\em Phys. Rev.} {\bf A 64}, (042315) (2001).

\bibitem{qsep}
S. Albeverio, S.M. Fei and D. Goswami, {\em Phys. Lett.} {\bf A},
91-96 (2001).

\bibitem{separ3}
S.M. Fei, X.H. Gao, X.H. Wang, Z.X. Wang and K. Wu,
{\em Phys. Lett. A} {\bf 300}, 559-566(2002).

\bibitem{HJ}
R.A. Horn and C.R. Johnson, {\it Matrix Analysis}, Cambridge University
Press, New York, 1985.

\bibitem{a8}
S.M. Fei and X.Q. Li,
{\it A Special Class of Matrices and Quantum Entanglement}, MIS-preprint 2002.

\end{thebibliography}
\end{document}